\documentclass{jfm}
\usepackage{graphicx}
\usepackage{epstopdf, epsfig}

\usepackage{xcolor}
\usepackage{amsmath}
\usepackage{adjustbox}
\usepackage{multirow}
\usepackage{tabularx, makecell, booktabs}
\usepackage{comment}

\shorttitle{Wave breaking with multilayer model}
\shortauthor{Jiarong Wu, Stéphane Popinet, and Luc Deike}

\title{Breaking wave field statistics with a multilayer model}

\author{Jiarong Wu\aff{1},
  Stéphane Popinet\aff{2}
 \and Luc Deike\aff{1,3}\corresp{\email{ldeike@princeton.edu}}}

\affiliation{\aff{1}Mechanical and Aerospace Engineering, Princeton University, Princeton, NJ, 08540
\aff{2}Institut Jean Le Rond d’Alembert, CNRS UMR 7190, Sorbonne Université, Paris 75005, France
\aff{3}High Meadows Environmental Institute, Princeton University, Princeton, NJ, 08540}
\begin{document}

\maketitle

\begin{abstract}
    The statistics of breaking wave fields is characterised within a novel multi-layer framework, which generalises the single-layer Saint-Venant system into a multi-layer and non-hydrostatic formulation of the Navier-Stokes equations. We simulate an ensemble of phase-resolved surface wave fields in physical space, where strong non-linearities including wave breaking are modelled, without surface overturning. We extract the kinematics of wave breaking by identifying breaking fronts and their speed, for freely evolving wave fields initialised with typical wind wave spectra. The $\Lambda(c)$ distribution, defined as the length of breaking fronts (per unit area) moving with speed $c$ to $c+dc$ following Phillips 1985, is reported for a broad range of conditions. We recover the $\Lambda(c) \propto c^{-6}$ scaling without any explicit wind forcing for steep enough wave fields. A scaling of $\Lambda(c)$ based solely on the mean square slope and peak wave phase speed is shown to describe the modelled breaking distributions well. The modelled breaking distributions are found to be in good agreement with field measurements and the proposed scaling is consistent with previous empirical formulations. The present work paves the way for simulations of the turbulent upper ocean directly coupled with realistic breaking waves dynamics, including Langmuir turbulence, and other sub-mesoscale processes.
\end{abstract}


\section{Introduction}
Wave breaking occurs at the ocean surface at moderate to high wind speed, with significant impacts on the transfer of momentum, energy, and mass between the ocean and the atmosphere \citep{MELVILLE1996,DEIKE2022}. When waves break, the water surface overturns, which generates sea spray and largely enhances the gas exchange. Visually it manifests as white-capping, widely observable at sea above a certain wind speed. Breaking acts as an energy sink for the waves: it limits the wave height by transferring the excessive wave energy into underwater turbulence and currents, therefore influencing the upper-ocean dynamics as well \citep{MCWILLIAMS2016,ROMERO2017}. 

Describing breaking waves analytically and numerically has been challenging due to its nonlinear nature and the fact that the interface becomes multi-valued. Considering a single breaker, scaling analysis have been successfully proposed for energy dissipation, validated by laboratory experiments \citep{DRAZEN2008,PERLIN2013}; and thanks to advances in numerical methods and increasing computational power, high fidelity simulations on single 3D breakers have emerged \citep{WANG2016,DEIKE2016,MOSTERT2022,GAO2021}. 



\citet{PHILLIPS1985} introduced the $\Lambda(\boldsymbol{c_b})$ distribution to describe the statistics of breaking waves, where $\Lambda(\boldsymbol{c_b})d\boldsymbol{c_b}$ is the expected length per unit sea surface area of breaking fronts propagating with speeds in the range of $(\boldsymbol{c_b}, \boldsymbol{c_b}+d\boldsymbol{c_b})$. The independent variable breaking front propagating speed $\boldsymbol{c_b}$ is chosen in place of wavenumber $\boldsymbol{k}$ because it is a more observable quantity. The link to the wave spectrum is made through the core assumption that $c_{b}$ is proportional to the wave phase speed $c$, which in turn relates to $k$ by the linear dispersion relation $c=\sqrt{gk}$. The omni-directional $\Lambda(c)$ distribution is predicted to have a $c^{-6}$ shape. The moments of the distribution have a physical interpretation, with the second moment related to the whitecap coverage, the third to mass exchange, the fourth to momentum flux and the fifth to energy dissipation by breaking \citep{PHILLIPS1985,KLEISS2010,ROMERO2019,DEIKE2018,DEIKE2022}.

Several observational studies have been conducted, which provide measurements of the $\Lambda(c)$ distribution, and its moments \citep{GEMMRICH2008,KLEISS2010,SUTHERLAND2013,BANNER2014,SCHWENDEMAN2015}, made possible by technical advancement including ship-borne and air-borne visible and infrared imagery. Scaling relations have been proposed to describe the breaking statistics for a wide range of conditions, but are facing the usual challenges in scatter of field data \citep{SUTHERLAND2013,DEIKE2018}, combined with ongoing discussions about the interpretation of \citet{PHILLIPS1985} original framework \citep{BANNER2014}. 


Beyond the single breaker description, numerical methods have so far been unable to describe the breaking statistics emerging from an ensemble of propagating surface waves. We propose a numerical framework, leveraging a novel multi-layer formulation of the Navier-Stokes equations and its numerical implementation \citep{POPINET2020}, which is able to capture the multi-scale nonlinear wave field, together with the intermittent incidences of breaking. The wave field is initialised using characteristic wind wave spectra based on field observations. We report the kinematics of the breaking statistics, $\Lambda(c)$, and its scaling with the mean-square slope and discuss how to link our results to field measurements. 


\section{Numerical method} \label{sec:multilayer} 
\subsection{The multi-layer framework}
\begin{figure}
    \centering
    \includegraphics[width=0.5\linewidth]{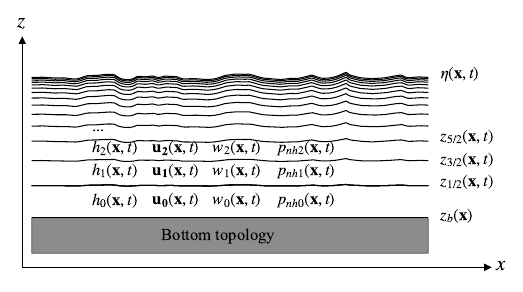}
    \caption{The layers in the multilayer model, and the fields of each layer. All the fields are functions of horizontal position $\boldsymbol{x}=(x,y)$ and time $t$. Due to the geometric progression choice, there is a fixed depth ratio between two adjacent layers.}
    \label{fig:layers_illustration}
\end{figure}

We introduce the modelling framework (sketched in figure \ref{fig:layers_illustration}) proposed by \citet{POPINET2020}, based on a vertically-Lagrangian discretisation of the Navier-Stokes equations. It extends the shallow-water single-layer Saint-Venant model to include multiple layers. 
We solve a weak form of the equation (vertically-integrated conservation laws) in a generalised vertical coordinate.  
Given $N_L$ layers in total, for layer number $l$ the mass and the momentum conservation equations are \citep{POPINET2020}:
\begin{eqnarray}
    \frac{\partial h_l}{\partial t} + \nabla_H \cdot (h\boldsymbol{u})_l &=& 0\label{eqn:numerical1} \\ 
    \frac{\partial (h\boldsymbol{u})_l}{\partial t} + \nabla_H \cdot (h\boldsymbol{u}\boldsymbol{u})_l &=& -g h_l \nabla_H \eta - \nabla_H(hp_{nh})_l + [p_{nh}\nabla_H z ]_l\label{eqn:numerical2} \\ 
    \frac{\partial (hw)_l}{\partial t} + \nabla_H \cdot(hw \boldsymbol{u})_l &=& -[p_{nh}]_l\label{eqn:numerical3} \\ 
    \nabla_H \cdot (h\boldsymbol{u})_l + [w-\boldsymbol{u}\cdot \nabla_H z ]_l &=& 0 \label{eqn:numerical4}
\end{eqnarray}
with $l$ the index of the layer, $h$ its thickness, $\boldsymbol{u}$, $w$ the horizontal and vertical components of the velocity, $p_{nh}$ the non-hydrostatic pressure (divided by the density). The surface elevation $\eta = z_b + \sum_{l=0}^{N_L} h_l$, and the $[\;]_l$ operator denotes the vertical difference, i.e. $[f]_l = f_{l+1/2} - f_{l-1/2}$. 
There are four unknowns $h_l$, $\boldsymbol{u_l}$, $w_l$ and $p_{nhl}$ for each layer. Equation \ref{eqn:numerical1} represents conservation of volume in each layer for layer thicknesses $h_l$ following material surfaces (i.e. the discretization is vertically Lagrangian). Equation \ref{eqn:numerical2} and \ref{eqn:numerical3} are the horizontal and vertical momentum equations. Equation \ref{eqn:numerical4} is the mass conservation equation. The time integration includes an `advection' step and a `remapping' step. In the `advection' step, equation \ref{eqn:numerical1} to \ref{eqn:numerical4} are advanced in time. In the `remapping' step, the layers are remapped, if necessary, onto a prescribed coordinate to prevent any severe distortion of the layer interface.

Note that this set of equations does not make any assumption on the slope of the layers, which explains the $\nabla_H z$ `metric' terms appearing in the horizontal momentum equation (\ref{eqn:numerical2}) and incompressibility condition (\ref{eqn:numerical4}). This is particularly important in the context of steep breaking waves. One can further demonstrate that this set of semi-discrete equations is a consistent discretisation of the incompressible Euler equations with a free-surface and bottom boundary \citep{POPINET2020}.
Note that, in the hydrostatic and small-slope limit, generalised vertical coordinates are widely used in ocean models \citep{GRIFFIES2020}, due to the anisotropic nature of geophysical flows. The choice of the target remapped discretisation is flexible and reflects physical considerations. Here, the remapping step uses a geometric progression of the layer thicknesses which ensures higher vertical resolution of the boundary layer under the free-surface.



The numerical schemes (spatial and temporal discretisations, field collocation, grid remapping, etc.) are described in detail in \citet{POPINET2020}, and ensure accurate dispersion relations and momentum conservation. 

\subsection{Numerical model for breaking}
The dissipation due to breaking is modelled with a simple, \emph{ad-hoc} approximation which can be related to the dissipation due to hydraulic jumps in the Saint-Venant system, known to be a surprisingly good first-order model for (shallow-water) breaking waves \citep{BROCCHINI2008}. The horizontal slope for any layer $\partial z/\partial x$ in equation \ref{eqn:numerical2} and \ref{eqn:numerical4} is limited by a maximum value $s_\text{max}$:
\begin{equation}
  \partial z/ \partial x = \left\{
    \begin{array}{ll}
      \partial z/\partial x, & |\partial z/\partial x| \leq s_\text{max}  \\[2pt]
      \text{sign}(\partial z/\partial x)s_\text{max}, & |\partial z/\partial x| > s_\text{max}.
  \end{array} \right.
\end{equation}
The maximum slope $s_{max}$ is set to be 0.577. The same applies to $\partial z/\partial y$. The slope limiter acts to stabilise the solver, and dissipates some amount of energy. 
We have tested that altering the value of $s_\text{max}$ between 0.4 and 0.6 does not change the numerical results significantly. Note that given enough horizontal resolution and vertical layers, and added viscous diffusion terms, the multilayer model converges to the full Navier-Stokes equations, with underwater turbulence, and the dissipation rate obtained from breaking is close to that obtained with direct numerical simulations. 

In the rest of the paper, we analyse the \emph{occurrence} of breaking fronts as geometric features of the surface height $\eta$, and investigate the relation between the wave statistics (wave spectrum) and breaking statistics (distribution of length of breaking crest).


\subsection{Numerical simulations of actively breaking wave fields} 
We initialise the wave field with an azimuth-integrated wavenumber spectrum of the following shape (inspired by field measurements such as \citet{ROMERO2010} and \citet{LENAIN2017}; see the discussion in \citet{DEIKE2022}),
\begin{equation}\label{eqn:spectrum_init}
\phi(k) = Pg^{-1/2}k^{-2.5}\exp[-1.25(k_p/k)^2].
\end{equation}
The value of $P$ controls how energetic the wave field is, and is of dimension of a velocity while $k_p$ is the peak wavenumber of the wave spectrum. Variations of the spectra parameters $k_p$ and $P$ can be summarised into a single non-dimensional global effective slope $k_p H_s$, where $H_s = 4\langle \eta^2 \rangle ^ {1/2}$ is the significant wave height. The global slope $k_p H_s$ is varied from 0.1 to 0.32 (almost no breaking waves to strongly breaking field).
The ratio $k_pL_0$, with $L_0$ the domain size, is kept constant at a sufficient large value ($k_pL_0=10\pi$) to avoid confinement effects, and we have verified that the results are independent from this ratio. The total water depth is chosen to be $2\pi/k_p$ to ensure a deep water condition.
The directional spectrum is $F(k,\theta) = (\phi(k)/k){\text{cos}^{N}(\theta)}/{\int_{-\pi/2}^{\pi/2} \cos^N(\theta) d\theta}$, with $\theta \in [-\pi/2, \pi/2]$.
The directional spreading is controlled by $N$, with $N=5$ for most cases, and we have tested $N=2$ (more spreading) and $N=10$ (less spreading). 

The initial wave field is a superposition of linear waves: $\eta = \sum_{i,j} a_{ij}\text{cos}(\psi_{ij})$, with the amplitude $a_{ij} = [2F(k_{xi},k_{yj})dk_xdk_y]^{1/2}$, and the initial random phase $\psi_{ij} = k_x x + k_y y + \psi_{\text{rand}}$. 
The corresponding orbital velocity is initialised similarly according to the linear wave relation. We use a uniformly spaced  initial grid of 32 $\times$ 33 array of ($k_{xi}$,$k_{yj}$). The wavenumbers are truncated, and chosen at discrete values of $k_{x} = ik_p/5$ for $i \in [1,32]$, and $k_{y} = jk_p/5$ for $j \in [-16,16]$, respectively. The horizontal resolution is $N_x = N_y= 1024$, and layer number $N_L = 15$, with a geometric progression common ratio 1.29. We have verified that the results presented here are numerically converged in terms of layer number (by running cases with 30 vertical layers); as well as horizontal resolution (see \S\ref{sec:Lambda_c}). 



\begin{figure}
    \centering
    \includegraphics[width=0.75\linewidth]{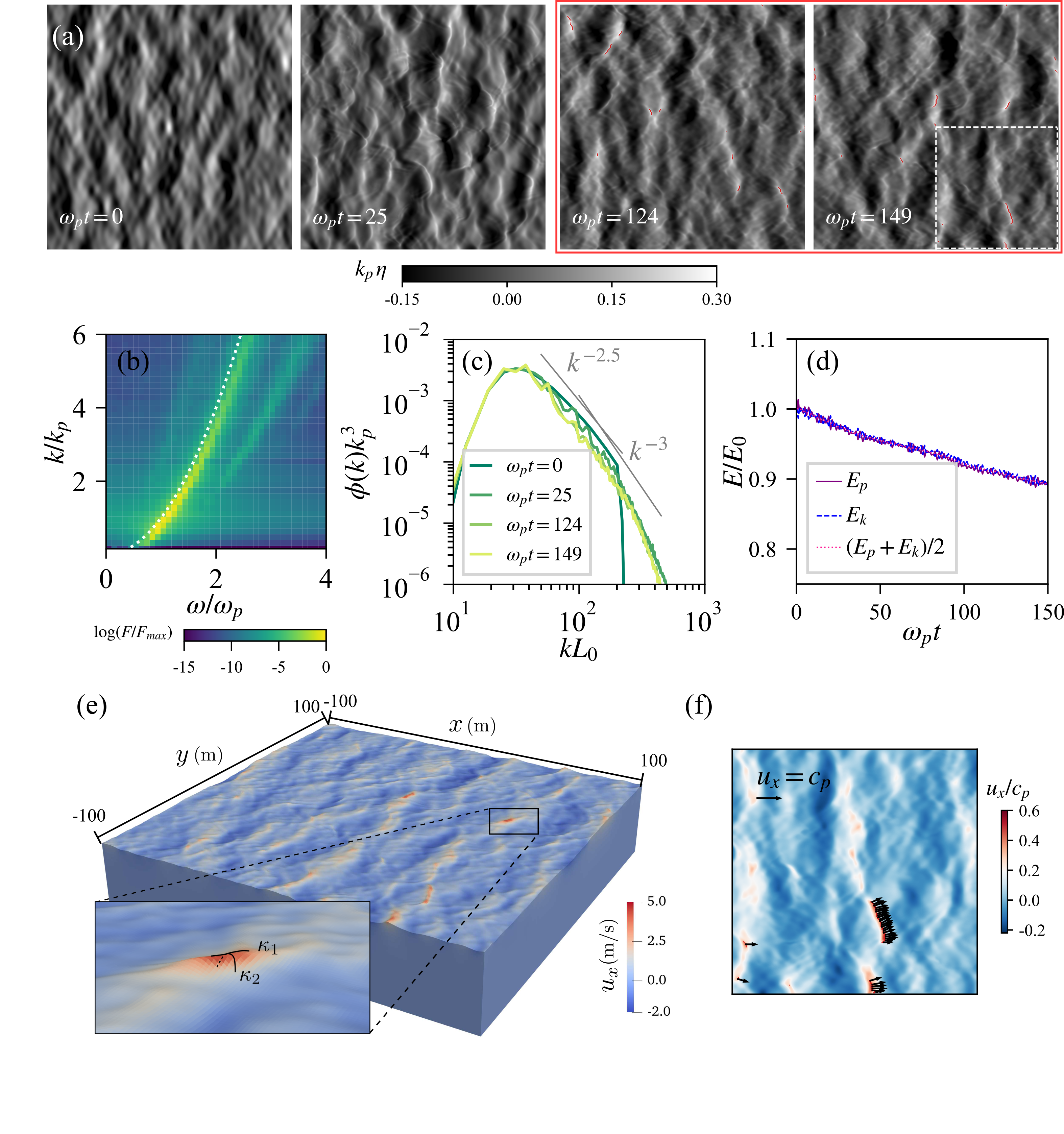}
    \label{fig:spectrum_evo_c} 
    \caption{(a) Snapshots of the wave field development for the case of effective slope $k_pH_s=0.233$. Breaking statistics are collected between $\omega_p t=124$ and $\omega_p t=149$ (indicated by the red box). (b) The wave energy spectrum on the frequency-wavenumber plane. The dotted white line is the linear dispersion relation of surface gravity waves $k=\omega^2/g$. (c) Time evolution of the omni-directional wave spectrum $\phi(k)$, corresponding to the snapshots in (a). (d) Energy evolution of the wave field. Purple: potential energy $E_p$; blue: kinetic energy $E_k$; pink: total energy. (e) 3D rendering of the breaking wave field with the colour indicating the surface layer flow velocity. Inset shows the curvature of the breaking fronts as the detection criterion. (f) A more focused view taken from the dotted white square in (a); The arrows are showing the velocity magnitude and direction of each length element of the breaking fronts.}
    \label{fig:spectrum_evo}
\end{figure}

Figure \ref{fig:spectrum_evo}(a) shows the time evolution of the wave field and the corresponding wave spectrum. The wave field is visually realistic. The space-time wave elevation spectrum is shown in figure \ref{fig:spectrum_evo}(b) and the energy is localised around the curve given by the gravity wave linear dispersion relation $\omega=\sqrt{gk}$, together with an extra branch corresponding to bound waves.
Since we start with a truncated spectrum, the initial wave field is smooth while the small scale features develop over time. There is an energy transfer into the higher wavenumbers which eventually leads to a stable spectrum shape. It is not a Kolmogorov--Zakharov type wave energy cascade as described by wave turbulence theory \citep{ZAKHAROV1992}, since the weak non-linearity assumption does not hold and  the small features are mostly generated by the breaking events.
A quasi-steady spectrum is obtained typically for $\omega_p t>100$ as seen on figure  \ref{fig:spectrum_evo}(c) with the wave statistics independent from the initial conditions, and we measure the breaking statistics between $\omega_p t=124$ and $\omega_p t=149$. Since there is no forcing mechanism, the total wave energy is slowly decaying (as shown in figure \ref{fig:spectrum_evo}(d)). The dissipation is primarily due to the slope-limiter, and is of the order of magnitude of known dissipation due to breaking \citep{DRAZEN2008}. We have also verified that the spatially and temporally-averaged statistics is a good representation of the ensemble average.



\subsection{Procedure of breaking front detection and velocity measurement}

The wave field evolves and breaking occurs intermittently in space and time. We detect the breaking fronts and their velocity, and construct the length of breaking crest distribution. The breaking fronts are defined geometrically as sharp enough ridges of the surface, as illustrated in figure \ref{fig:spectrum_evo}(e).  Given a surface elevation $\eta(x,y)$ at one time instance, we find its Gaussian curvature $\kappa_1$ and $\kappa_2$, and determine the location of the breaking fronts by the threshold $\kappa_2 < -3k_p$ (`ridges' of the $\eta$ surface), which works well across the different scales.
After the breaking regions (areas) are detected, we extract the breaking fronts (lines), shown in figure \ref{fig:spectrum_evo}(f). Then we use the surface layer Eulerian velocity ($\boldsymbol{u_{l-1}}$ in figure \ref{fig:layers_illustration}) as an estimate of the Lagrangian velocity of the breaking fronts $\boldsymbol{c_b}$. The velocity is mapped on each discretised cell on the lines, which represents an element of length $L_0/N_x$. Figure \ref{fig:spectrum_evo}(f) shows the mapped velocity magnitude and direction with arrows. The directionality of $\boldsymbol{c_b}$ is not discussed in this work, i.e. we only consider the magnitude $c_b=|\boldsymbol{c_b}|$. We have tested an alternative velocity mapping method by computing the correlation function between two consecutive images of the detected crests (similar to particle tracking velocimetry), and found no significant difference in the velocity magnitude detected or the resulting $\Lambda(c_b)$ distribution. 



We follow \citet{PHILLIPS1985,KLEISS2010,SUTHERLAND2013} and assume that $c=c_b$; and we use a correspondence between the breaking front velocity and the underlying wavenumber through the dispersion relation $c=\sqrt{g/k}$. 
We note that observations have shown that $c_{b} = \alpha c$ where $\alpha$ is between 0.7 to 0.95 \citep{RAPP1990,BANNER2014,ROMERO2019}, at least for large breakers. In the processing we filter out the smaller scale breakers by imposing a filter $\eta(x,y) > 2.5 \langle \eta \rangle^{1/2}$. It means that only the large breakers with surface elevation above 2.5 rms value are included. As a result, no further corrections for the underlying long wave orbital velocity is needed.



\section{Statistics of wave breaking} \label{sec:Lambda_c} 
\subsection{Wave statistics}

We study the relation of the breaking statistics with the wave spectrum. Figure \ref{fig:spectra+lambdac}(a) shows the non-dimensional wave spectra for the various conditions, with variations in spectrum maxima larger than one order of magnitude, and described by power laws ranging from $\phi(k)\propto k^{-2.5}$ to $\phi(k)\propto k^{-3}$. Although the energy close to the peak frequency varies, a fixed level of saturation seems reached for the steeper cases with overlapping spectra in the $k^{-3}$ range.


Together with the global slope $k_pH_s$, wave statistics can be characterised by the root mean square slope $\sigma$ \citep{MUNK2009}, which is more sensitive to high frequencies. The low-pass filtered steepness parameter $\mu(k)$ is defined as the cumulative root mean square slope: $\mu^2(k) = \int_0^k k'^2\phi(k') dk'$, and $\sigma$ is the asymptotic value of $\mu$ with a cutoff at the highest wavenumber we can numerically resolve $k_{max}$: $\sigma^2 = \mu^2(k\to k_{max})$. As we see in figure \ref{fig:spectra+lambdac}(a), the value of $\mu(k)$ plateaus due to the drop-off of the spectrum. In weak nonlinear theories (such as wave-turbulence theory), $\mu < 0.1$ is used to justify the asymptotic expansions, at least for the range of $k$ considered \citep{ZAKHAROV1992}. All the breaking cases in our simulation have $\mu$ closer or higher than 0.1 underlying the strong non-linearities of the breaking wave field. The correlation between the two global slope parameters $k_pH_s$ (zeroth moment of the spectrum) and $\sigma$ (second moment of the spectrum) is shown by figure \ref{fig:spectra+lambdac}(b), which we caution is specific to the spectrum shape.



\subsection{$\Lambda(c)$ distribution}

\begin{figure} 
\centering
\includegraphics[width=0.75\linewidth]{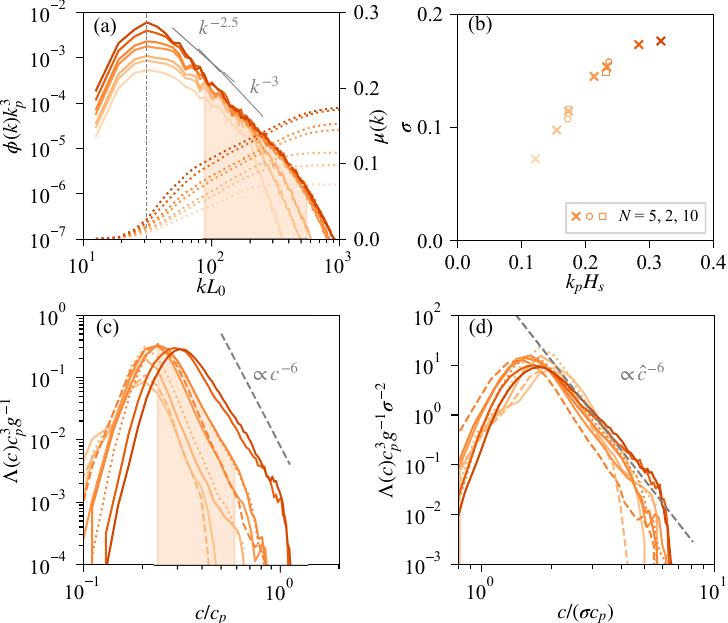}
\caption{\label{fig:spectra+lambdac} (a) The wave energy spectra in non-dimensional form; the vertical gray line is  $k_pL_0 = 10\pi$. Darker colour indicates larger global slope $k_pH_s$ (see (b) for the values). (b) The correlation of root mean square slope $\sigma$ and the global slope $k_pH_s$ in the simulated cases. (c) The non-dimensional breaking distribution $\Lambda(c)$ normalised by $c_p$ and $g$. Solid lines: directional spreading parameter $N=5$; dashed lines: $N=2$; dotted lines: $N=10$. (d) Proposed scaling for the $\Lambda(c)$ distribution using $\sigma$ and $c_p$.}
\end{figure}

Figure \ref{fig:spectra+lambdac}(c) shows the breaking distribution $\Lambda(c)$ for increasing $k_p H_s$ (and $\sigma$) values and the various directionalities. 
There is a clear peak indicating the most probable smaller breakers, increasing from $c= 0.2c_p$ to $0.3c_p$ when the slope increases. There is no breaking for the smallest $\sigma=0.065$ case ($k_p H_s= 0.117$) (not shown in figure \ref{fig:spectra+lambdac}(c)) an increase in slope to $\sigma=0.085$ ($k_p H_s=0.150$) and $\sigma=0.101$ ($k_p H_s= 0.169$) starts to generate breakers. The extent of breaking speeds is further increased for the steeper cases with $\sigma>0.101$, with a clear $\Lambda(c)\propto c^{-6}$ scaling up to around $0.9c_p$. It indicates that there exists a critical value of $\sigma$, below which the breaking wave field is not saturated, with the threshold expected to depend on the spectrum shape.

The shaded area in figure \ref{fig:spectra+lambdac}(c) spans the range of the breaker velocity between the peak $\Lambda_{max}$ and  $\Lambda(c)=0.01\Lambda_{max}$ (for the case of $\sigma=0.153$), where a $\Lambda(c)\propto c^{-6}$ scaling can be clearly observed. The same range in the $k$-space is shaded in figure \ref{fig:spectra+lambdac}(a) as well. The upper limit of $\Lambda(c)=0.01\Lambda_{max}$ corresponds to a lower limit of $k\approx4k_p$. Above that velocity, breakers near $k_p$ are very rare. We note that removing the filter of $\eta(x,y) > 2.5 \langle \eta \rangle^{1/2}$ only changes the part of the $\Lambda(c)$ distribution left of the peak, but does not affect the part with $c$ larger than the peak. Similarly, further increasing the horizontal resolution would extend the move up and toward even smaller $c$, but the presented $\Lambda(c)\propto c^{-6}$ is unchanged.

$\Lambda(c)$ distributions from spectra of different directionality $N$ are also shown with different lines (dashed lines indicate more spreading ($N=2$) and dotted lines less spreading ($N=10$)). For steep enough cases ($\sigma > 0.101 $), there is little difference in the $\Lambda(c)$ distribution between cases with different $N$, while for intermediate steepness ($\sigma = 0.85$ and $0.101$), there is a notable sensitivity to $N$. For the $N=10$ cases with more concentrated wave energy, there is overall more breaking events.

Non-dimensional scalings of the $\Lambda(c)$ distribution have been proposed, with \citet{PHILLIPS1985} using only the wind speed, while papers based on field data used a combination of wind speed, wave spectrum peak speed and significant wave height \citep{SUTHERLAND2013,DEIKE2018}. Since we have no wind forcing in the simulations, the breaking statistics is expected to scale only with the non-dimensional slope $\sigma$ and spectrum peak speed $c_p$. By rescaling $c$ using $\hat{c} = c/(\sigma c_p)$ and $\Lambda(c)$ using $\hat{\Lambda}(c) = \Lambda(c)c_p^3g^{-1}\sigma^{-2}$, we obtain a normalised distribution:
\begin{equation}
    \Lambda(c)c_p^3g^{-1}\sigma^{-2} \propto \left(c/(\sigma c_p)\right)^{-6} \label{eqn:lambdac_scaling}
\end{equation}
shown in figure \ref{fig:spectra+lambdac}(d), which collapses not only the $\hat{c}^{-6}$ power law region but also the peak location well (for the steep enough cases). 
Alternatively, since the effective slope $k_pH_s$ and $\sigma$ are correlated, a scaling using $(k_pH_s)^{1/2}$ could be proposed. However, we have found that $\sigma$ works better than $k_pH_s$ as a scaling parameter in this case.

\subsection{Comparison to the \citet{PHILLIPS1985} theory and observations}

\citet{PHILLIPS1985} predicted a purely wind-based scaling $\Lambda(c)\propto u_*^3 g c^{-6}$ through an energy balance argument. The wave action balance equation $d[g\phi(k)/\omega]/dt = S_{nl}(k)+S_{in}(k)+S_{diss}(k)$ involves the following source terms: divergence of the nonlinear energy flux $S_{nl}$, wind input $S_{in}$, and dissipation due to breaking $S_{diss}$, written as \citep{PHILLIPS1985}
\begin{equation}
    S_{nl} \propto gk^{-3}B^3(k), \; S_{in} \propto gk^{-3}(\frac{u_*}{c})^2 B(k), \; \text{and} \; S_{diss} \propto gk^{-3} f(B(k))
\end{equation}
with the saturation $B(k)=k^3\phi(k)$, and $f(B(k))$ a functional dependence solely on $B(k)$ (assuming that breaking and consequent dissipation `\textit{are the result of local excesses, however these excesses are produced}'). The balance between $S_{nl}$ and $S_{diss}$ leads to $f(B)\propto B^3$, and therefore $S_{diss} \propto gk^{-3}B^{-3}$.
The breaking front distribution $\Lambda(c)$ is then obtained by writing the equality between dissipation in the $k$-space and the $c$-space: $\epsilon(k)dk = \epsilon(c)dc$. The LHS is $\epsilon(k)dk = (S_{diss}\omega) dk$; the RHS can be related to the fifth moment of $\Lambda(c)$ through a scaling argument $\epsilon(c)dc = bg^{-1}c^5\Lambda(c)dc$, where $b$ is a non-dimensional breaking parameter \citep{DUNCAN1981,PHILLIPS1985}. Substituting a spectral shape of $\phi(k)\propto k^{-5/2}$ into the $S_{diss}$ would then lead to $\Lambda(c)\propto c^{-6}$. Considering the equilibrium range, $S_{nl} \propto S_{in}$ \citep{PHILLIPS1985}, gives $\phi(k) \propto u_*g^{-1/2}k^{-2.5}$, which leads $\Lambda(c) \propto u_*^3gc^{-6}$.

Several field campaigns have observed the $c^{-6}$ power-law, despite also finding that the purely wind-based prefactor $u_*^3$ does not describe the data well. Empirical modifications have been proposed \citep{SUTHERLAND2013,DEIKE2018} using $\sqrt{gH_s}$ and $c_p$ in addition to $u_*$, in the form of $\Lambda(c)c_p^3 g^{-1} (c_p/u_*)^{1/2}\propto (c/\sqrt{g H_s})^{-6}$, which significantly improved the collapse between data sets.

\begin{figure}
\centering
\includegraphics[height=0.27\linewidth]{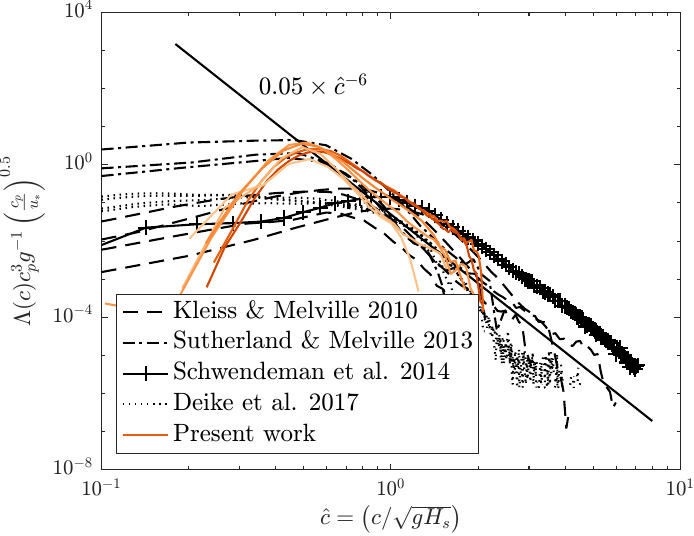}
\includegraphics[height=0.27\linewidth]{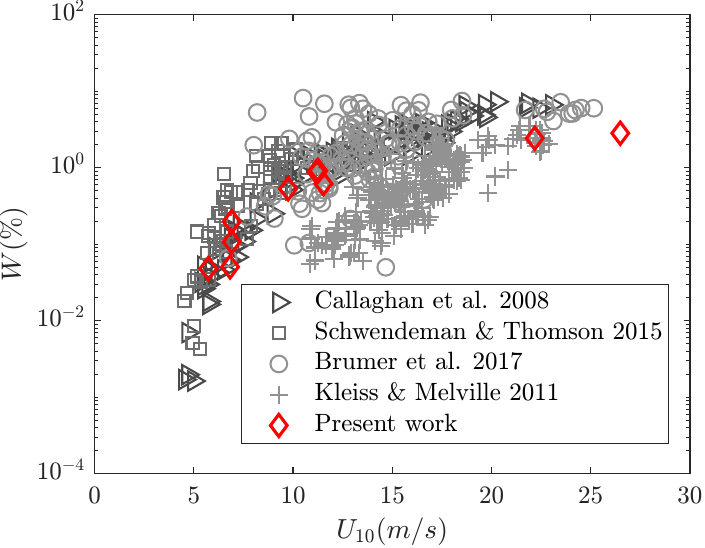}
\caption{\label{fig:data} Comparison with observational data. (a) Rescaled $\Lambda(c)$ distribution following \cite{SUTHERLAND2013} with simplifications \citep{DEIKE2022}. (b) Whitecap coverage $W$ as a function of 10-meter wind speed $U_{10}$. 
}
\end{figure}

To better interpret the empirical scaling found in field observations, we perform re-analysis of the numerical data together with the field data. 
We examine the slope-based scaling in the present work and the mixed-wind-slope-based scaling found in field data. Since there is no explicit wind forcing in our simulations, the information of wind speed and fetch/duration are encoded in the spectrum. We use the empirical (but very robust) fetch-limited relationships \citep{TOBA1972}, that link the non-dimensional wave energy $gH_s u_*^{-2}$ and the non-dimensional frequency $\omega_p g^{-1}u_*$ (wave age $u_*/c_p$) by
\begin{equation}
gH_s/u_*^2 = C (u_*/c_p)^{-3/2} \label{eqn:fetch_limited}
\end{equation}
where $C$ is an order 1 constant. Using (\ref{eqn:fetch_limited}) it is straightforward to show that the scaling (\ref{eqn:lambdac_scaling}) with $\sigma \propto \sqrt{k_pH_s}$ is equivalent to the scaling from \citet{SUTHERLAND2013}, since $k_pH_s \propto (c_p/u_*)^{1/2}$. Figure \ref{fig:data}(a) shows the breaking distributions scaled following \citet{SUTHERLAND2013} (the wind speed $u_*$ is inferred using (\ref{eqn:fetch_limited}) in our modelled cases), and good agreement is observed, indicating that the slope-based scaling is fully compatible with the mixed-wind-slope-based scalings in the field. Note that the observational data are obtained from complex sea states, and do not necessarily have the same spectrum shape as the current numerical data. 




The wave field at a certain time and space is the result of the wind forcing history, and breaking is caused by excessive energy in the wave spectrum. For a mature and steep enough wave field, breaking (particularly those at large scale) is primarily dictated by the wave spectrum itself, while for younger and less steep wave fields, breaking can be more closely coupled to wind forcing. It explains why the slope-based scaling or a mixed-wind-slope-based scaling can better fit data from various sea states.

Finally, we can infer classic breaking metrics such as the whitecap coverage from our simulations and compare with more field data sets. The whitecap coverage $W$ quantifies the fraction of the wave surface covered by white foam, and can be estimated through the second moment of $\Lambda(c)$ as $W=2\pi\gamma g^{-1}\int c^2\Lambda(c) \:d{c}$, where $\gamma$ is a dimensionless constant representing the ratio of breaking time to wave period (here $\gamma = 0.56$ following \citet{ROMERO2019}). Figure \ref{fig:data}(b) shows $W$ as a function of the 10-meter wind speed $U_{10}$ (estimated from $u_*$ for our data using the COARE parameterisation \citep{EDSON2013}). The modelled whitecap coverage falls within the scatter of recent data sets \citep{CALLAGHAN2008,KLEISS2010,SCHWENDEMAN2015,BRUMER2017}.

\section{Conclusion}

We demonstrate that a novel multilayer model \citep{POPINET2020} can be used to study the breaking statistics associated with an ensemble of phase-resolved surface waves simulated in the physical space. We analyse the breaking front distribution introduced by \cite{PHILLIPS1985}, and find good agreement with field observations. The breaking distribution follows $\Lambda(c) \propto c^{-6}$ even in the absence of wind input, and can be scaled by the mean square slope, indicating that the universal breaking kinematics is primarily governed by the wave field itself, while the wind controls the development of the wave spectrum. The proposed scaling in terms of the mean square slope is fully compatible with empirical relationships used to describe field data. 




Our approach provides an unprecedented numerical framework to study breaking statistics for complex wave spectra, which could help to understand the breaking distribution in complex seas (in the presence of swell or currents) and complement existing modelling approaches such as \citet{ROMERO2019}. In addition to the physical discussion of breaking statistics, we demonstrate the capability of the multi-layer approach to solve highly nonlinear geophysical flows with strong vertical-horizontal anisotropy.

\bibliographystyle{jfm}
\bibliography{ref}

\end{document}